\documentstyle[12pt]{article}
\begin{document}
\begin{center}

{\large \bf General Relativistic Effects on Quantum
            Interference and the Principle of Equivalence}\\

\vspace{8mm}
      K.K. Nandi$^{~ a,c,}$\footnote{E-mail address: kamalnandi@hotmail.com}
   and Yuan-Zhong Zhang$^{~ b,c,}$\footnote{E-mail address: yzhang@itp.ac.cn} \\
\vspace{4mm}
  {\footnotesize{\it
  $^a$Department of Mathematics, University of North Bengal,
      Darjeeling (W.B.) 734430, India\\
  $^b$CCAST(World Lab.), P. O. Box 8730, Beijing 100080, China \\
  $^c$Institute of Theoretical Physics, Chinese Academy of Sciences,
          P.O. Box 2735,\\ Beijing 100080, China}}\\
\end{center}

\vspace{8mm}

\begin{abstract}
 Using a novel approach, we work out the general relativistic effects
 on the quantum interference of de Broglie waves associated with thermal
 neutrons. The unified general formula is consistent with special
 relativistic results in the flat space limit. It is also shown that
 the exact geodesic equation contains in a natural way a gravitational
 analog of the Aharonov-Bohm effect. We work out two examples, one in
 general relativity and the other in heterotic string theory, in order
 to obtain the first order gravitational  correction terms to the quantum
 fringe shift. Measurement of these terms is closely related to the
 validity of the equivalence principle at a quantum level.\\

\noindent
  PACS number(s): 95.85.Ry, 12.15.Ff, 14.60.Lm
\end{abstract}

\vspace{10mm}
 \section{Introduction}
Recently, there has been a revival of interest in the topic
dealing with the terrestrial test of the equivalence principle of
Einstein's general relativity at a classical level [1]. The works
focus mainly on the phenomena of various types of couplings such
as the classical spin-spin, spin-orbital or spin-gravitational
couplings [2]. On the other hand, investigations of the
gravitational effects on the quantum phenomena are expected to
provide informations that could be useful not only for testing the
equivalence principle at a quantum level but also for the
development of a full theory of quantum gravity [3]. For instance,
a semiclassical treatment reveals that quantum uncertainty in the
source variables induces uncertainties in the metric components of
gravity in a specific manner [4]. Another important phenomenon is
the neutrino flavor oscillation induced by gravity [5-8] or by the
recoil from virtual D-branes [9]. A different kind of theoretical
approach in the calculations of the phase of quantum particles,
neutrinos included, has yielded a very interesting result: The
Dirac spin 1/2 particle has the exact covariant Stodolsky [10]
phase $S/\hbar  = (1/\hbar )\int {p_\mu  } dx^\mu $ in a static
gravity field [11]. Some of these investigations could provide
appropriate theoretical backgrounds in the atmospheric neutrino
experiments [12], or in the observations involving $\gamma$-ray
bursts [13].

Recent interests also include the study of quantum interference
fringes for thermal neutrons traveling along two different paths
[14,15]. The interference with itself of the de Broglie wave
associated with an ensemble of particles allows us to predict, via
Huygen's principle, the motion of the wave. The presence of an
external field modifies the motion and useful information about it
could be gathered if one knows how the external field causes a
shift in the interference fringes. A classic example is the
Aharonov-Bohm (AB) fringe shift which provides information as to
how the motion of electrons are modified in the presence of a
magnetic potential [16]. The AB effect predicts the same gauge
invariant fringe shift, viz.
 $ \frac{e}{{\hbar c_0 }}\oint {A_\mu  } dx^\mu $
 between the interfering beams both in special relativistic
and nonrelativistic theories, where $A_\mu$ is the electromagnetic
4-potential, $e$ is the electronic charge, $\hbar$ is Planck's
constant and $c_0$ is the light speed in vacuum. However,
 the situation becomes more complicated, both experimentally
 and theoretically, when one introduces gravity and rotation.
On the experimental side, the developments in the matterwave
interferometry have shown a greater promise (over photon
 interferometry) in the measurements of the influences on
the fringe shifts due to Earth's gravity and rotation. The neutron
interferometry of the earlier experiments by Colella, Overhauser
and Werner (COW experiment)[17] and an improved configuration of
the later experiment by Werner, Staudenmann and Colella (WSC
experiment) [18] provide information about the effects of Earth's
gravity and Coriolis force on the fringe shift at a classical,
that is, non-general relativistic level.

On the theoretical side, a special relativistic treatment of the
quantum fringe shift for thermal neutrons has been proposed by
Anandan [15]. However, the effects of gravity and rotation are
still considered as separate components while using special
relativistic expressions for energy and momentum. On the other
hand, a more comprehensive analysis involving gravity and inertial
forces requires the unified framework of general relativity that
combines those forces into a single geometrical framework. The
special relativistic effects should then follow as a limiting
case. The current literature seems to still lack such a unified
approach. This has been noted in a recent investigation by Zhang
and Beesham [14]. However, they consider only Schwarzschild
gravity and consequently the relativistic or the Coriolis parts do
not appear in their formulation.

In this paper, adopting a new approach developed in a series of
recent papers [19], we wish to work out the exact general
relativistic equation of the quantum fringe shift for thermal
neutrons. A ``Coriolis" force appears {\it naturally} in the exact
form of the geodesic equation. This gives rise to a gravitational
analog of the quantum AB effect. The curved spacetime
contributions to the special relativistic and Coriolis effects on
the fringe shift are displayed. As an immediate corollary, it is
argued that the measurements of these contributions would
establish the validity or otherwise of the equivalence principle
at a quantum level. For illustrations, we calculate the first
order gravitational contributions coming from the Kerr metric of
general relativity and the Kerr-Sen metric of heterotic string
theory [20]. The latter will enable us to also envisage the string
effects.

The paper is organized as follows: In Sec.II, for ready reference,
Anandan's special relativistic (SR) approach is summarized. In
order to be reasonably exhaustive, we enumerate, in Sec.III, the
salient features of the present approach leading to an appropriate
form of the geodesic equation. In Sec.IV, the general relativistic
equation for the quantum fringe shift is proposed. Two examples
are considered in Sec.V, while some concluding remarks are given
in Sec.VI.

\section{Special relativistic quantum phase shift}
In the literature, there exist several deductions of the
individual parts of the special relativistic contributions to
quantum phase shift [10,15,21,22]. However, using intuitive
arguments, Anandan [15] proposed a general formula for the quantum
phase shift based on a correspondence between the fringe shift and
the classical special relativistic equations of motion. This
remarkable general formula at once gives rise to all the above
individual components. Consider two de Broglie wavelets
originating at A and interfering at B, one traveling along ADB and
the other along ACB, the plane ADBC being fixed to Earth and
aligned vertically (Fig.1). The vertical direction is determined
by the resultant of gravity and the centrifugal force of Earth. In
the absence of any external forces, BD=BC and we take CD = $d$ ,
and AB = $l$. The external field causes a shift in the position of
B. Let $ \tilde v $ be the velocity of the classical particle and
$\delta \tilde v$ be the change in the tranverse velocity in the
plane of interference. Then,
 $$\Delta \phi _{SR}  = \tilde \kappa d\frac{{\delta \tilde
v}}{{\tilde v}} = \frac{{\tilde \kappa d}}{{\tilde
v}}\frac{{d\tilde v_ \bot  }}{{dt}}\frac{l}{{\tilde v}} =
\frac{{\tilde pA}}{{\hbar \tilde v^2 }}\frac{{d\tilde v_ \bot
}}{{dt}} = \frac{{\tilde \gamma mAd\tilde v_ \bot  /dt}}{{\tilde
v\hbar}},                                         \eqno(1)$$
 where $\tilde\gamma = \left({1-\tilde v^2 /c_0^2}\right)^{-1/2}$,
  $A=ld$  is the planar area (assumed small) enclosed by the two
paths of interfering beams, $m$ is the neutron mass,
 $\tilde p( =\hbar \tilde \kappa )$ is the momentum in flat space,
  and $d\tilde v_ \bot  /dt$ is the transverse component
of acceleration in the vertical direction in the plane ADBC. (We
have made a slight change in the notation, that is, added a tilde
for flat space quantities, in order to be consistent with our
later equations.) The formula (1) exhibits a general
correspondence between the quantum phase shift and the classical
equation of motion.

The following cases were considered: (i) A particle with charge
$e$ moving in a magnetic field $B$.
 Then, $\tilde \gamma d\tilde v_ \bot  /dt = eB\tilde v/m$, which,
 when used in Eq.(1), immediately yields the AB effect, viz.,
 $$\Delta \phi _{AB}  = eBA/\hbar.              \eqno(2)$$
(ii) A spinless particle of mass $m$ in a gravitational field,
$d\tilde v_ \bot  /dt =  - g $ where $g$ is the gravitational
acceleration on the surface of the Earth. Using the expression for
the relativistic momentum $\tilde \gamma ^2  = 1 + \tilde p^2 /m^2
c_0^2 $ and the de Broglie relation $\tilde p =\hbar\tilde \kappa
$, one obtains from Eq.(1), the following result:
  $$\Delta \phi _g  =  - gAm^2 /\hbar ^2 \tilde \kappa  - gA\tilde
  \kappa /c_0^2.                                   \eqno(3)$$
The COW experiment was accurate enough to measure the first
quantum term, but not the second, that is, the so called
relativistic term. It may be mentioned here that the term
 $ -gAm^2 /\hbar ^2 \tilde \kappa $ could also be derived by an
analysis involving the motion of neutrons along radial and cross
radial directions [14]. However, the advantage of Anandan's
approach, which we adopt in this paper, is that it offers a
simpler alternative and yet leads to more results than just this
term, as we can see in this section. (iii) A particle moving in
the Coriolis force field of Earth so that, to first order in
$\Omega $, $ d\tilde{v}_{\bot}/dt = 2\left| \vec{\Omega}\times
\vec{\tilde{v}}\right| = 2\Omega_n \tilde{v} $, where $\Omega_n $
is the component of Earth's angular velocity $\vec \Omega$ normal
to the apparatus. Using the Planck-Einstein law $E/c_0  = \hbar
\tilde \omega  = mc_0 \tilde \gamma $, one finds from Eq.(1) that
  $$\Delta \phi_{cor} = \frac{2 \tilde \omega A \Omega_n}
      {c_0}                                         \eqno(4)$$
The WSC experiment has tested this effect to within a good
accuracy. (iv) This effect comes from the coupling of particle's
spin to the background curvature. Again using (1) together with
the Mathisson-Papapetrou force [23], the shift comes out to be
 $$\Delta \phi _{s.c.}  =  - \frac{{\hbar GMA\tilde \omega
   }}{{mc_0^3 R^3 }},                        \eqno(5)$$
where $G$ is the Newtonian gravitational constant, $M$ is Earth's
mass and $R$ is the distance from the center. This effect is too
tiny to be measurable at present and we do not discuss it here.
With all the above in view, we proceed to familiarize the readers
with the salient features of our approach in the next section.

\section{The approach: Basic equations}
The basic idea of the method is to introduce the idea of an
optical-mechanical analogy in general relativity [19]. The analogy
provides an excellent tool that enables one to visualize the
problems of geometrical optics as problems of classical mechanics
and vice versa. The first step is to find out an optical
refractive index $n$ that is formally equivalent to the
geometrized gravity field. This step in itself is not new. In the
study of optical propagation in a gravity field, this index has
been used in the literature {\it albeit} in an approximate form.
For instance, the index equivalent to the exterior Schwarzschild
field is usually taken to be $n \approx 1 + 2MG/rc_0^2$, where $M$
is the central gravitating mass. In our approach, however, we
consider the exact expression shown below.

Consider a static, spherically symmetric, but not necessarily
vacuum, solution of general relativity written in isotropic
coordinates
 $$ds^2  = h(\vec r)c_0^2 dt^2  -\Phi ^{ - 2}(\vec r)\left|{d\vec
   r} \right|^2,                                \eqno(6)$$
where $\vec r \equiv (x,y,z)$ or $(r,\theta ,\varphi)$, and
  $h, \Phi$ could be the solution of Einstein's field equations.
Many metrics of physical interest can be put into this isotropic
form. The coordinate speed of light $c(\vec r)$ is determined by
the condition that the geodesic be null $\left(ds^2 =0\right)$:
 $$c(\vec r) = \left| {\frac{{d\vec r}}{{dt}}} \right| = c_0 \Phi
  \sqrt h,                                       \eqno(7)$$
which immediately provides an effective index of refraction for
light in the gravitational field given by
 $$n(\vec r) = \frac{1}{{\Phi \sqrt h }}.       \eqno(8)$$
As the next step, we point out that the concept of the optical
mechanical analogy can be used to recast the geodesic equation for
{\it both} massive and massless particles into a single, exact
Newtonian ``F=ma" type of equation given by [19]
 $$\frac{{d^2 \vec r}}{{dA^2 }} = \vec \nabla \left( {\frac{1}{2}N^2
c_0^2 } \right),N(\vec r) = n(\vec r)\sqrt {1 - \frac{{m^2 c_0^4
h}}{{E_0^2 }}} ,dA = \frac{{dt}}{{n^2 }},        \eqno(9)$$
 where $m$ is the rest mass of the particle, $E_0$ is the conserved
total energy, $\vec \nabla $ is the gradient operator. All the
standard geodesic equations in Schwarzschild gravity including
some new insights in cosmology follow from the above equation
[19]. This remarkably simple feature of the geodesic equations is
brought about by the use of the stepping parameter $A$, first
introduced by Evans and Rosenquist [24]. Eqs. (9) provide an easy
way to introduce quantum relations so that the geodesic motion of
a particle can be interpreted as motion of de Broglie matter waves
in a dispersive medium with an effective index of refraction
$N(\vec r,\lambda ) = n(\vec r)/\sqrt {1 + (\lambda /\lambda _c
)^2 }$ where $\lambda_c$ is the Compton wavelength given by
$\lambda _c  = 2\pi \hbar /mc_0$ and $\lambda  = \tilde \lambda
\Phi ^{ - 1}$ is the physical wavelength measured in a gravity
field. This interpretation allows us to extend in a
straightforward manner the classical optical-mechanical analogy
into the quantum regime.

Alsing [25] has subsequently extended the method to broader class
of metrics in general relativity and this work is going to provide
the basic foundation of what follows. Consider the most general
form of the metric given by
 $$ds^2  = g_{00} c_0^2 dt^2  + 2g_{0i} c_0 dtdx^i  + g_{ij} dx^i
  dx^j ,\quad \quad i,j = 1,2,3,                    \eqno(10)$$
Define the proper time $d\tau$ , proper length $dl$  and the
velocity $v$ measured with respect to this proper time as
 $$v^i  = \frac{{dx^i }}{{d\tau }},                 \eqno(11)$$
 $$\left. {\frac{{ds}}{{c_0 }}} \right|_{v = 0}  = d\tau  =
  \frac{{\sqrt h }}{{c_0 }}(c_0 dt - g_i dx^i ),
  \quad \quad g_{00}=h,                            \eqno(12)$$
 $$dl^2=\gamma_{ij}dx^i dx^j =\left({-g_{ij}+\frac{{g_{0i}g_{0j}}}
    {{g_{00} }}} \right)dx^i dx^j, \quad \quad
    v^2  = \gamma _{ij} v^i v^j,                    \eqno(13)$$
 $$g_i =-\frac{{g_{0i}}}{{g_{00}}},\quad \quad g^i =\gamma^{ij}
    g_j  =  - g^{0i}.                               \eqno(14)$$
The $g_i$ and the proper velocity  $v_i$ are vectors defined in
the 3-space characterized by the metric $\gamma_{ij}$, which is
used to raise or lower the indices of these 3-vectors. Now, the
metric (10) can be rewritten as
 $$ds^2  = h(c_0 dt-g_i dx^i)^2 \left( {1 - v^2 /c_0^2 }\right).
                                                     \eqno(15)$$
The conserved energy $E$ is given by
 $$E = mc_0^2 g_{0\alpha }\frac{{dx^\alpha}}{{ds}}=\frac{{mc_0^2
  \sqrt h }}{{\sqrt {1 - v^2 /c_0^2 } }}, \quad\quad
     \alpha =0, 1, 2, 3.            \eqno(16)$$
     The variational principle for the
geodesics following from Eq.(15) is given by
 $$\delta \int\limits_{\vec x_1 ,t_1 }^{\vec x_2 ,t_2 }{mc_0 ds =}
\delta \int\limits_{\vec x_1 ,t_1 }^{\vec x_2 ,t_2 } {Ldt = }
\delta \int\limits_{\vec x_1 ,t_1 }^{\vec x_2 ,t_2 } {mc_0^2 \sqrt
{h(\vec r)} \sqrt {1 - v^2 (\vec r,\tilde v)/c_0^2 } } \tilde
\beta (\vec r,\tilde v)dt = 0,                     \eqno(17)$$
where
 $$\tilde \beta  =1-\frac{{g_i \tilde v^i }}{{c_0 }},\quad \quad
\tilde v^i  = \frac{{dx^i }}{{dt}},\quad \quad v^i  =
\frac{{\tilde v^i }}{{\tilde \beta \sqrt h }}.       \eqno(18)$$
Henceforth, unless specifically restored, we take $G = c_0  = 1$.
From the Lagrangian $L$, let us find the momenta conjugate to
$\tilde v^i$. This is given by
 $$\frac{{\partial L}}{{\partial \tilde v^i }} = E\left( {g_i  +
   \frac{{v_i }}{{\sqrt h }}} \right).              \eqno(19)$$
Since the vectors $g_i$ and $v_i$ are defined in a space with the
metric $\gamma_{ij}$, we can identify the right hand side of the
Eq.(19) as a vector in the same 3-space. Let us call it the
momentum 3-vector
 $$p_i  \equiv E\left( {g_i  + \frac{{v_i }}{{\sqrt h }}}\right).
                                                   \eqno(20)$$
It can also be verified that $H = \frac{{\partial L}}{{\partial
\tilde v^i }}\tilde v^i  - L \equiv p_i \tilde v^i  - L = E$,
which is a constant along the trajectory of a particle as stated
in Eq.(16). Hence, it is possible to introduce Maupertuis
principle $\delta \int\limits_{\vec x_1 }^{\vec x_2 } {p_i \tilde
v^i } dt = 0$. Assuming further that the spatial part of the
metric could be written in an isotropic form $dl = dl_E /\Phi
,dl_E  = \delta _{ij} dx^i dx^j ,\gamma _{ij}  = \delta _{ij} \Phi
^{ - 2}$, we get  $v^2  = n^2 \tilde v^2 /\tilde \beta ^2$, and
the Maupertuis variational principle yields, after some
manipulations introducing the parameter $A$, the geodesic equation
in the form
 $$\frac{{d^2 \vec r}}{{dA^2 }} = \vec \nabla \left( {\frac{{n^2 v^2
  }}{2}} \right) + \frac{{d\vec r}}{{dA}} \times (\vec \nabla
  \times \vec g),                                   \eqno(21)$$
 $$dA = n^{ - 2} \tilde \beta dt,                   \eqno(22)$$
where $\vec g \equiv (g_i )$. This Newtonian form of the geodesic
equation, valid for both massless and massive particles, has been
obtained under the only assumption that the spatial part of the
metric could be written in an isotropic form. On eliminating $A$
from Eqs.(21) and (22), we can find the rotational contributions
to the well known Schwarzschild orbits for light and planets [25].

Equation similar to Eq.(21) has actually been derived in the
literature [26], but only under the assumptions of small velocity
and weak gravity. On the other hand, the novelty of the above set
of equations [Eq.(21), (22)] is that they are {\it exact} and that
they hold for {\it all velocities as well as in strong gravity
fields}. The Eq.(21) admits an immediate interpretation as
describing the motion of a particle in a ``potential" $( - 1/2)n^2
v^2$ and subjected to a ``Coriolis" force $\frac{{d\vec r}}{{dA}}
\times (\vec \nabla  \times \vec g)$, which would appear, for
instance, in the absence of gravity $(n=1, h=1)$ in a coordinate
system rotating with angular velocity
$\mathord{\buildrel{\lower3pt\hbox{$\scriptscriptstyle\rightharpoonup$}}
\over \Omega }  = (1/2)(\vec \nabla  \times \vec g)$. Another
advantage is that Eq.(21) is expressed in terms of the velocity
measured in proper time in a rotating field and this is exactly
the velocity of the quantum particle measured by an observer
comoving with the interferometer.

\section{General relativistic effect on the fringe shift}
Our approach indicates that it is possible to simulate the quantum
interference in Earth's gravity as an experiment in a rotating
medium described by $n$ and other quantities that take care of the
nonlinearities. In order to find a general relativistic version of
Eq.(1), we should express the relevant quantities in terms of
their proper values, recalling that in a gravity field, one is
able to measure only proper, and not coordinate, quantities. We
make the reasonable assumption that $n, h$  and $\Phi$ do not vary
appreciably over the dimensions of the apparatus (a few square
centimeters). From Eq.(12), we get
 $$d\tau  = \tilde \beta \sqrt h dt,                  \eqno(23)$$
and defining $d^2 \vec r/d\tau ^2  \equiv d\vec v_ \bot  /d\tau$,
we can write, using Eqs.(21) and (22), the transverse component of
the proper acceleration as
 $$\frac{{d\vec v_ \bot  }}{{d\tau }} = \left( {\frac{1}{{2n^2 h}}}
\right)\vec \nabla \left( {v^2 } \right) + \left( {\frac{2}{{n^2
\sqrt h }}} \right)\left( {\vec v \times \vec \Omega } \right),
                                                      \eqno(24)$$
where $\vec \Omega  = \frac{1}{2}\left( {\vec \nabla  \times \vec
g} \right)$. Further, noting that  $p^2  = p_i p^i  = \gamma _{ij}
p^i p^j$, and introducing quantum relations, we have, from
Eqs.(20) and (16), respectively,
 $$p = \left[ {\frac{{m\sqrt h }}{{\sqrt {1 - v^2 } }}} \right]\left[
{\frac{{v^2 }}{h} + \frac{{2g_i v^i }}{{\sqrt h }} + g_i g^i }
\right]^{\frac{1}{2}}  = \hbar \kappa,                 \eqno(25)$$
 $$E = \frac{{m\sqrt h }}{{\sqrt {1 - v^2 } }} = \hbar \omega.
                                                     \eqno(26)$$
Reasoning along the lines similar to those in Ref.[15], and using
Eqs.(24) and (25), we can write, under the assumption stated
above, the general relativistic version of Eq.(1) as
 $$\Delta \phi _{GR}  = \left( {\frac{{dv_ \bot  }}{{d\tau }}}
\right)\left( {\frac{{pA}}{{\hbar v^2 }}} \right) = \left[ {\left(
{\frac{1}{{2n^2 h}}} \right)\frac{d}{{dr}}\left( {v^2 } \right) +
\left( {\frac{2}{{n^2 \sqrt h }}} \right)\left| {\vec v \times
\vec \Omega } \right|} \right]\left( {\frac{{pA}}{{\hbar v^2 }}}
\right) = \Delta \phi _1  + \Delta \phi _2,        \eqno(27)$$
 in which $dv_ \bot  /d\tau ( \equiv d^2 r/d\tau ^2  - rd\varphi
/d\tau )$ is the radial component of acceleration. The rotational
effects are represented by $g_i$ (through $\tilde \beta $), which
occur in $v^i$ [Eq.(18)], $p$ [Eq.(25)] and, of course, in the
Coriolis term $\left| {\vec v \times \vec \Omega } \right|$.
Strictly speaking, even the area  $A$ above would no longer be
planar in the gravity field. But considering the miniscule
dimension of the apparatus in a weak gravity field on the surface
of the Earth, we may disregard its departure from planarity in the
practical computations. The equation (27) for the quantum fringe
shift is the main proposition in our paper.

If one considers, as is done in Ref.[15], that gravity is given by
the Newtonian law $\tilde v^2 =\left|{d\vec r/dt}\right|^2 =2M/r$
in an otherwise flat space $(n=1, h=1)$ and that the rotational
effects are contained exclusively in the classical Coriolis term
so that $\tilde \beta  = 1$, then one has
 $$\Delta \phi _{SR}  = \left[-g + 2\left|\vec{\tilde{v}}\times
\vec{\Omega} \right| \right]\left(\frac{\tilde{p}A}{\hbar
 \tilde{v}^2} \right),                            \eqno(28)$$
  where
 $$ - g = \frac{1}{2}\frac{d}{{dr}}(\tilde v^2 ) = M/R^2.
                                                    \eqno(29)$$
One can see how beautifully the effects of gravity and rotation,
usually considered as separate components in the literature [cases
(ii) and (iii) in Sec.2] , are synthesized into a single equation.

In order to further explore the nature of Eq.(21), we note that
Sakurai [27] presented an interesting derivation of the phase
shift induced by Earth's rotation given by Eq.(4). It shows that
the effect of Coriolis force can be transcribed as an AB-type
effect. This transcription finds a natural place in our approach.
It was already stated in the previous section that Eq.(21)
exhibits a ``Coriolis" type force. The equation can alternatively
be regarded as a ``Lorentz" force equation for the motion of a
classical particle with charge to mass ratio $q/m$ in a static
classical electromagnetic field. By scaling the stepping parameter
$A$ to a variable having the dimension of time $t'$, and
identifying the scalar potential $\psi (\vec r) = (1/2)n^2 v^2 ( -
q/m)^{ - 1}$ and the vector potential $\vec A(\vec r) = \vec
g(q/m)^{ - 1}$, we have from the four potential  $A^\mu   = (\psi
,\vec A)$, the electric and magnetic fields as
 $$\vec E =  - \frac{{\partial \vec A}}{{\partial t'}} - \vec \nabla
\psi ,\quad \quad \vec B = \vec \nabla  \times \vec A. \eqno(30)$$
The last equation can be rewritten as $q\vec B = 2m\vec \Omega$,
where $\vec \Omega  = (1/2)(\vec \nabla  \times \vec g)$ is the
rotational velocity of the Earth with respect to distant stars
[see Eq.(34) in the next section]. Now, it is well known that the
AB phase difference in the interference region is proportional to
the magnetic flux enclosed by the two paths as in Fig.1 and is
given by the line integral over the two routes:
 $$\phi _{ADB}- \phi _{ACB}  = \left( {q/\hbar }\right)\oint{\vec
A.d\vec l = \left( {q/\hbar } \right)} \int {\vec B\cdot d\vec
\sigma}.                                         \eqno(31)$$
Putting the value for $q\vec B$ in the above, we get $\phi _{ADB}
- \phi _{ACB}  = \left( {2m/\hbar } \right)\int {\vec \Omega \cdot
d\vec \sigma }$, where $d\vec \sigma$ is the oriented area $A$ of
the parallelogram ADBC. Using the Planck-Einstein law, we get just
the expression (4), but with the difference that our $\vec \Omega$
is completely determined by the metric 3-vector $\vec g$. It thus
follows that the gravitational analog of the AB fringe shift is
already embedded into the exact Eq.(21), thereby strengthening the
weak field, low velocity analysis in Ref. [26]. One may also
regard the fringe shift due to the Coriolis term as somewhat
analogous to the quantized version of the Sagnac effect without
the particle mass term [15].

\input{epsf.sty}
\begin{figure}
\begin{center}
  \leavevmode
 \epsfysize=1.8in
  \epsfbox{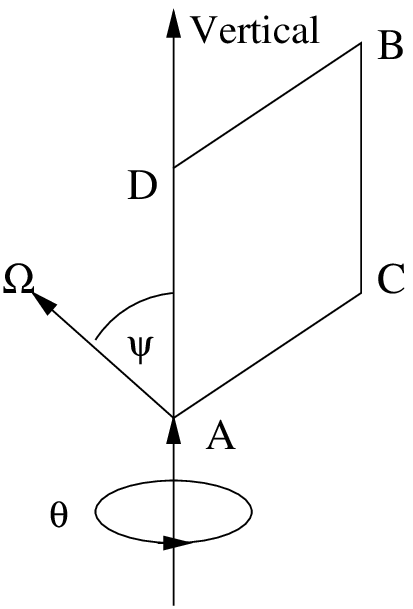}
\end{center}
 {\footnotesize Fig. 1.  Experiment to measure the phase shift due
to earth's rotation in neutron interference. The interfering beams
travel along paths ADB and ACB in a neutron interferometer. $\vec
\Omega$ is the earth's angular velocity: $\Omega _n  = \Omega \sin
\psi \sin \theta. $ }
\end{figure}

\section{Examples}
(a) \underline{The Kerr solution}

We shall first consider the Kerr solution although, probably, it
does not adequately represent any realistic rotating source due to
quadrupole considerations. Nonetheless, we can have an idea as to
how the corrections appear in different terms. Since the apparatus
is situated on the surface of the Earth, we can take the metric to
lowest order in $M/r$ and fixing the apparatus on the equator
($\psi  = \pi /2 $, Fig.1), we have:
 $$\left. {d\tau ^2 } \right|_{\theta  = \pi /2,M/r <  < 1}  \cong
\left( {1 - \frac{{2M}}{r}} \right)\left[ {dt +
\frac{{2Ma}}{r}d\varphi } \right]^2  - \left( {1 + \frac{{2M}}{r}}
\right)\left[ {dr^2  + r^2 d\varphi ^2 } \right],  \eqno(32)$$
where $r$ is the isotropic radial variable, $J \equiv Ma$ is the
Earth's angular momentum. Then one has the following expressions:
 $$\Phi  \cong 1 - M/r,h \cong 1 - 2M/r,n \cong 1 + 2M/r,g_\varphi
 \cong  - 2Ma/r,                                \eqno(33)$$
 $$\left| {\vec \nabla  \times \vec g} \right| = r^{ - 1}
\frac{{dg_\varphi  }}{{dr}} = \frac{{2Ma}}{{r^3 }} = 2\left|
{\mathord{\buildrel{\lower3pt\hbox{$\scriptscriptstyle\rightharpoonup$}}
\over \Omega } } \right|.                       \eqno(34)$$
Assuming the velocity of the particle to be $v_0$ in the
asymptotic region where $h=1$, we have $E/m = E' = 1/\sqrt {1 -
v_0^2}$, and using the expression for $E$ [Eq.(16)], we find
 $$v^2  = 1 - h(1 - v_0^2 ).                   \eqno(35)$$
Since, to first order, $h \cong 1 - 2M/r$, we have, on Earth's
surface,
 $$\left( {1/2} \right)d(v^2 )/dr =  - g\left( {1 - v_0^2 } \right).
                                                \eqno(36)$$
Here we encounter a situation that reminds us of the statements of
Cardall and Fuller [28]: It is not possible to extract the pure
gravitational field $g$ in the locally inertial frame; there also
appears a contribution $gv^2_{0}$  in Eq.(36) coming in from the
asymptotic region. However, under the low velocity approximation,
we can disregard this extra term for practical computations and
take the usual Newtonian equation (29) so that
 $$v \cong \tilde v = \sqrt {2M/r}.             \eqno(37)$$

Let us assess the terms, to lowest order, appearing in the second
square bracket in Eq.(25). An analysis of the orbits of Eq.(21)
suggests that [25], $\tilde v^\varphi   = d\varphi /dt \cong
2Ma/r^3$, so that we have
 $$\tilde \beta  = 1 - g_i \tilde v^i  = 1 - g_\varphi  \tilde
  v^\varphi   \cong 1 + 4J^2 /r^4,               \eqno(38)$$
 $$\frac{{2g_i v^i }}{{\sqrt h }} \cong 2\left( {1 + \frac{M}{r}}
\right)\left[ {\frac{{g_\varphi  \tilde v^\varphi  }}{{\tilde
\beta }}} \right] \cong 8\left( {1 + \frac{M}{r}}
\right)\frac{{J^2 }}{{r^4 }},                     \eqno(39)$$
 $$g_i g^i  = g_\varphi  g^\varphi   = \left( {\frac{{2Ma}}{r}}
\right)g^{0\varphi }  \cong \frac{{4J^2 }}{{r^4 }}. \eqno(40)$$
Therefore, to first order $(1/r)$, the last two terms in the
second square bracket in the expression for $p$ do not contribute,
and hence, we may write, using Eq.(37): $p \cong \tilde p =
\frac{{m\tilde v}}{{\sqrt {1 - \tilde v^2 } }} = \hbar \tilde
\kappa$. Together with the expression $hn^2  = \Phi ^{ - 2}  \cong
1 + 2M/r$, and using the identity, $1 + \frac{{\tilde p^2 }}{{m^2
}} = \frac{1}{{1 - \tilde v^2 }}$, we get from Eq.(27), the
effects of only static gravity in terms of $\Phi$:
 $$\Delta \phi _1  = \Phi ^2 \left( { - \frac{{gAm^2 }}{{\hbar ^2
\tilde \kappa }}} \right) - \Phi ^2 \left( {gA\tilde \kappa /c_0^2
} \right).                                   \eqno(41)$$
 The rotational effect is then contained only in the Coriolis term
$\left| {\vec v \times \vec \Omega } \right|$ with $\vec \Omega$
determined by the metric coefficients. Writing $E = \hbar \tilde
\omega \sqrt h$ and using it in $\Delta \phi _2$, we find that the
term is modified also by the same factor
 $$\Delta \phi _2  = \Phi ^2 (2A\tilde \omega \Omega _n /c_0 ).
                                                \eqno(42)$$
Using the approximate dispersion relation
 $$\tilde \omega ^2  - \tilde \kappa ^2  = \left( {\frac{{m^2
}}{{\hbar ^2 }}} \right)\left( {\frac{1}{{1 - \tilde v^2 }}}
\right)\left( {h - \tilde v^2 } \right),        \eqno(43)$$
 we get
 $$\Delta \phi _2 =\Phi ^2 \left( {\frac{{2\Omega _n Am}}{\hbar}}
\right)+{\rm small~ relativistic ~ terms } ~ O(c_0^{ - 2}).
                                           \eqno(44)$$
Note that velocity dependent dispersion relations similar to
Eq.(43), or more generally to the one following from the Eqs.(25)
and (26), occur also in other works, although in different
contexts [9,29].  With the expression for $\Phi$ given in Eq.(33),
we easily see the first order corrections to respective special
relativistic effects. The importance of the correction factors
will be pointed out in the last section.

(b) \underline{The string theory}

Low energy effective field theory describing heterotic string
theory also produces rotating black hole solutions. A classical
exact solution has been found by Sen [20] which we refer to here
as the Kerr-Sen solution. The Kerr-Sen metric in the Einstein
frame closely resembles the familiar Kerr solution in
Boyer-Lindquist coordinates [30]:
 $$d\tau ^2  = \left( {1 - \frac{{2M}}{\rho }} \right)dt^2  - \Sigma
\left( {\frac{{d\rho ^2 }}{\Delta } + d\theta ^2 } \right) -
\left[ {\rho \left( {\rho  + \xi } \right) + a^2  + \frac{{2M\rho
a^2 \sin ^2 \theta }}{\Sigma }} \right]\sin {}^2\theta d\varphi ^2
$$
 $$ + \frac{{4M\rho a\sin ^2 \theta }}{\Sigma }dtd\varphi,
                           \eqno(45)$$
where the dilatonic field $\Psi$, the electromagnetic potentials
$A_{t}, A_{\phi}$ and the tensor gauge potential $B_{t\phi}$  are
given by
 $$\Psi  =  - \ln \left[ {\frac{\Sigma }{{\rho ^2  + a^2 \cos ^2
\theta }}} \right], ~~ A_\varphi   =  - \frac{{2\sqrt 2 a\rho
Q\sin ^2 \theta }}{\Sigma }, ~~ A_t  = \frac{{2\sqrt 2 \rho
Q}}{\Sigma }, ~~ B_{t\varphi }= \frac{{a\rho Q^2 \sin ^2 \theta
}}{{M\Sigma }},$$
 $$ \Sigma  = \rho (\rho  + \xi ) + a^2 \cos ^2 \theta ,\quad \Delta
= \rho (\rho  + \xi ) + a^2  - 2M\rho ,\quad \xi  = Q^2 /M.
                                  \eqno(46)$$
The metric describes a black hole with mass $M$, dilatonic charge
$Q$, angular momentum $aM$, and magnetic dipole moment $aQ$ The
quantity $\xi$ has the dimension of length. For $Q=0$ the metric
reduces to the Kerr solution of general relativity and for $a=0$
it reduces to the Gibbons-Garfinkle-Horowitz-Strominger (GGHS)
black hole solution [31] with the redefinition $\rho  \to \rho  -
\xi$.

Under the weak field approximation as in (a), we have, in the
equatorial plane,
 $$\left. {d\tau ^2 } \right|_{\theta  = \pi /2,M/(\rho  + \xi ) <  <
1}  \cong \left( {1 - \frac{{2M}}{{\rho  + \xi }}} \right)\left[
{dt + \frac{{2Ma}}{{\rho  + \xi }}d\varphi } \right]^2  - \left(
{1 + \frac{{2M}}{{\rho  + \xi }}} \right)d\rho ^2  - \rho (\rho  +
\xi )d\varphi ^2.                             \eqno(47)$$

Under the reasonable assumption that $\xi  <  < \rho$, we can
write $\rho (\rho  + \xi ) \cong (\rho  + \xi )^2$. In this case,
we can adopt the isotropic radial variable $r$ related to the
standard variable $\rho$ via $r = (\rho  + \xi )e^{ - M/(\rho  +
\xi )}$ and retaining terms in lowest order in $M/(\rho  + \xi )$,
we have $r \cong \rho  + \xi  - M.$  The metric for large $r$
then has the form
 $$\left. {d\tau ^2 } \right|_{\theta  = \pi /2,M/(r + \xi ) <  < 1}
\cong \left( {1 - \frac{{2M}}{{r + \xi }}} \right)\left[ {dt +
\frac{{2Ma}}{{r + \xi }}d\varphi } \right]^2  - \left( {1 +
\frac{{2M}}{{r + \xi }}} \right)\left[dr^2  + (r + \xi )^2
d\varphi ^2 \right].                          \eqno(48)$$ This has
exactly the same form as Eq.(32) under the identification $r \to r
+ \xi$ and all the calculations in (a) can be carried out in a
similar manner. We have
 $$\Phi ^2 \cong 1 - \frac{{2M}}{r}+ \frac{{2M(\xi  + M)}}{{r^2 }},
                                              \eqno(49)$$
which indicates that the effect of the dilaton $\xi$ appears only
in the order $O\left(r^{-2}\right)$  and its measurement appears
to be beyond the present capabilities.

\section{Concluding remarks}
Our analysis above is primarily intended to fill a gap in the
literature pointed out recently by Zhang and Beesham [14]. The
proposed Eq.(27), together with inputs from Eqs.(24)-(26),
constitute the desired unified, covariant version of the quantum
fringe shift for thermal neutrons. Generally, the equations
exhibit both the effects of gravity and rotation in an inseparable
manner. Apart from this, the present work has a bearing on the
(weak) principle of equivalence. If a viable theory of quantum
gravity has to be formulated, the validity of this principle must
be tested even at a quantum mechanical level. Anandan [15]
suggested a method which involves carrying out the experiment with
the interferometer at a distance $r$ from the axis of rotation of
a turntable. If the principle is valid quantum mechanically, one
should obtain the phase shift Eq.(3) with $g=\Omega^2 r$. On the
other hand from the classical point of view, it is well known that
the principle is embedded in Einstein's theory of general
relativity in the form of geodesic equations which follow from the
Bianchi identities. Hence, a natural question arises if we can
write the geodesic equation in a form that at once reveals the
effect of gravity and rotation on the quantum fringe shift while,
at the same time, leads to corresponding special relativistic
results in the limit. Our analysis above shows that it is indeed
possible and a unified expression for the general relativistic
quantum fringe shift was proposed. Thus, a test of the principle
in the quantum regime involves looking for general relativistic
correction terms in an interferometric experiment. It does no
longer seem necessary to regard the equivalence principle as a
mere equality of Newtonian forces, viz., $g=\Omega^2 r$, since
general relativity encapsules something more in its nonlinearities
than just this. As an aside, an interesting analogy between the
Coriolis and the AB effect has been pointed out.

We considered two examples, one from general relativity and the
other from string theory. With an eye to practical feasibilities,
we had worked out only the first order corrections, viz.,
$\frac{{2MG}}{{Rc_0^2 }}\left( {\frac{{gAm^2 }}{{\hbar ^2 \tilde
\kappa }}} \right),\quad \frac{{2MG}}{{Rc_0^2 }}\left(
{\frac{{gA\tilde \kappa }}{{c_0^2 }}} \right),$ and
$\frac{{2MG}}{{Rc_0^2 }}\left( {\frac{{2\Omega _n Am}}{\hbar }}
\right)$ to the respective special relativistic terms measured in
the COW or WSC-type experiments. In this approximation, one has to
be content with the metric rotational contribution being
represented only by the angular velocity part $\Omega_n$. This is
a bit unfortunate but seems inescapable. The corrections are quite
consistent from an intuitive point of view as well, although they
arose here from a reasonably exact equation (27). Since, on the
surface of the Earth, $2MG/Rc_0^2  \approx 10^{ - 9}$,
measurements of these corrections in an Earth bound configuration
require a sensitivity at least of the order of $10^9$ times more
than what the present setup offers [32]. The string contributions
appear only in the second order in the form $2Q^2/R^2$ and
multiplying it with special relativistic terms, one obtains
extremely tiny corrections that look fairly out of question at
present [30].

As a final comment, we wish to point out that the validity of the
principle of equivalence for {\it charged} particles still appears
to be a debatable issue [33]. At a fundamental level, the issue
relates to the question as to how the (radiation) energy is to be
defined: Is it a conserved quantity associated with Lorentz boosts
in Minkowski space [34] or with the Killing symmetry of time
translations [33]? The resolution of this question has a deep
relevance to whether the breakdown of Lorentz invariance could be
interpreted as a violation of the principle of equivalence,
although an affirmative interpretation is commonly adopted in the
literature [9,35]. These issues will be addressed in a future
communication.

{\bf Acknowledgments}
 One of us (KKN) wishes to thank the Director,
Professor Ouyang Zhong Can, for providing hospitality and
excellent working facilities at ITP, CAS. Useful discussions with
Professor Han Ying Guo, Professor Ming Yu, both at ITP, and Dr. A.
Bhadra are gratefully acknowledged. Sun Liqun is thanked for
technical assistance. The work was in part supported by the
TWAS-UNESCO visiting associateship program of ICTP, Italy, and
also by NNSFC under Grant Nos. 10175070 and 10047004, as well as
by NKBRSF G19990754.

\end{document}